# Challenges and perspectives in computational deconvolution of genomics data

Lana X. Garmire[1$], Yijun Li[2], Qianhui Huang[1], Chuan Xu[3], Sarah Teichmann[3], Naftali Kaminski[4], Matteo Pellegrini[5], Quan Nguyen[6], Andrew E. Teschendorff[7,8]

1: Department of Computational Medicine and Bioinformatics, University of Michigan, Ann Arbor, MI, USA
2: Department of Biostatistics, University of Michigan, Ann Arbor, MI, USA
3: Wellcome Sanger Institute, Wellcome Genome Campus, Hinxton, Cambridge, UK
4: Pulmonary, Critical Care & Sleep Medicine, Yale University School of Medicine, New Haven, CT. USA
5: Molecular, Cell and Developmental Biology, University of California Los Angeles, Los Angeles, CA, USA
6: Curtin Medical School, Curtin Health Innovation Research Institute, Curtin University, Perth, WA 6102, Australia
7: CAS Key Laboratory of Computational Biology, Shanghai Institute of Nutrition and Health, University of Chinese Academy of Sciences, Chinese Academy of Sciences, Shanghai, 200031, China
8: UCL Cancer Institute, University College London, London, WC1E 8BT, UK

$: corresponding author Email: lgarmire@med.umich.edu

## Abstract
Deciphering cell type heterogeneity is crucial for systematically understanding tissue homeostasis and its dysregulation in diseases. Computational deconvolution is an efficient approach estimating cell type abundances from a variety of omics data. Despite significant methodological progress in computational deconvolution in recent years, challenges are still outstanding. Here we enlist four significant challenges related to computational deconvolution, from the quality of the reference data, generation of ground truth data, limitations of computational methodologies, and benchmarking design and implementation. Finally, we make recommendations on reference data generation, new directions of computational methodologies and strategies to promote rigorous benchmarking.

## Introduction

The biology of any complex tissue is directly dependent on cells, the basic units of biological functions. There are estimated to be over 200 different types of cells in the human body[1]. Virtually all tissues in multicellular organisms are heterogeneous, composed of multiple cell types. Thus, cell type heterogeneity is crucial to consider across many biomedical domains. For example, cell type heterogeneity has gained increasing attention in cancer treatment. The state of the tumor microenvironment, including the cell types, their proportions and interactions with tumor cells can have a significant impact on treatment efficacy, metastasis, and survival[2]. Therefore, deciphering cell type heterogeneity is crucial for systematically understanding homeostasis under healthy conditions and the dysregulation in diseases. On the other hand, when conducting translational research, ignoring confounders due to variations in cell type abundances often leads to biased or even erroneous scientific conclusions in the downstream analyses, such as differential gene expression or differential methylation[3,4]. Thus, adjusting cell-type heterogeneity is not only necessary but a critical step to ensure unbiased identification of disease signatures.

Experimental approaches to decipher each tissue are expensive and time consuming, limited to certain types of cells, and subjective to impurity even within detected cell types. To overcome these issues, the alternative computational process called "cell-type deconvolution" has emerged as an important area of research in the field of genomics. Mathematically, the problem of computational deconvolution can be formulated as $E = S * C$. E is the matrix of bulk-tissue level feature representation, and can be modeled by multiplying a reference matrix S indicating cell-type-specific features by a cell type proportion matrix C. Such a generalized matrix factorization procedure can be solved either by deterministic linear models, probabilistic models or deep learning approaches. Many genomics data types will benefit from such computational advancements, such as gene expression, epigenetics (eg. DNA methylation and ATAC-Seq[5,6]), and spatial omics. Though the underlying computing principles are similar across different omics, we focus on three main omics applications of computational deconvolution methods based on the type data in the matrix E (**Figure 1)**: bulk-tissue gene expression, DNA methylation, or non-single-cell (e.g., 'spot'-based) spatial transcriptomics data [7,8]. For readers interested in deconvolution methods for other omics, eg. DeconPeaker[9], DC3[10] for ATAC-Seq, please refer to other studies.

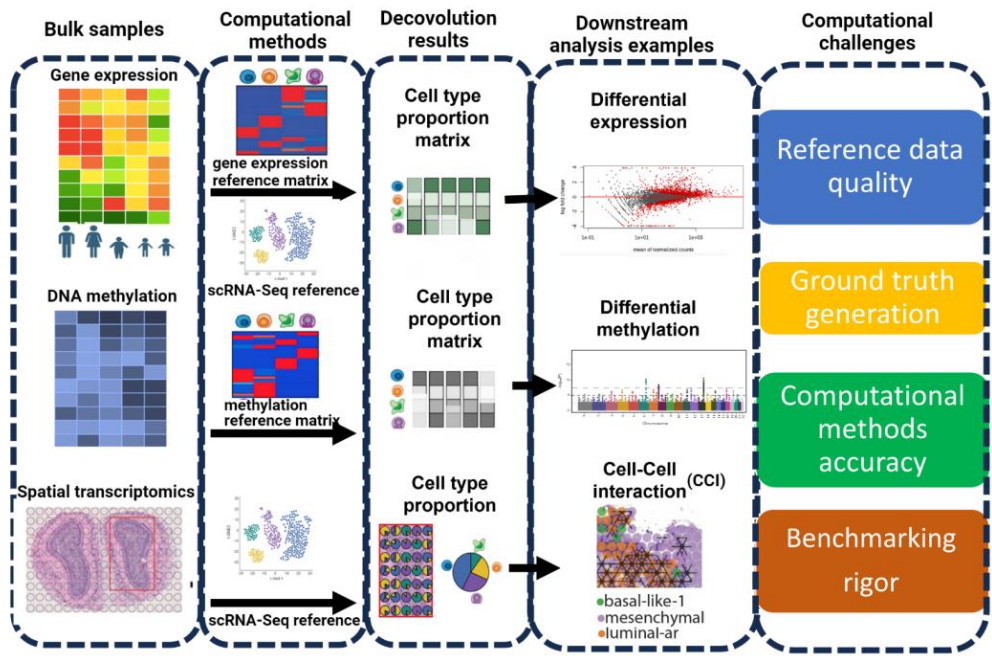

*Figure 1. Overview of computational deconvolution in various genomics data types and related challenges. Deconvolution of bulk-tissue transcriptomic data can be categorized into bulk-tissue cell-type specific reference-based and single cell RNA-Seq reference-based approaches; Deconvolution of bulk-tissue DNA methylation data is predominantly based on bulk-tissue cell-type specific references; Deconvolution of non-single cell resolution spatial transcriptomics data utilize the single cell RNA-Seq data as the reference. The outputs for bulk-tissue transcriptomic and DNA methylation data are cell-type proportion matrices. The outputs for spatial transcriptomics data are represented by spatially resolved pie charts. Colors in each pie chart correspond to the proportions of different types of cells in each "Spot" of the spatial transcriptomics data. Deconvolution results significantly impact downstream analyses, exemplified by differential expression (for bulk-tissue transcriptomics data), differential methylation aka. epigenome-wide association analysis (for bulk-tissue DNA methylation data) and cell-cell interaction analysis (for spatial transcriptomcis data). Four computation-related challenges stand out: reference data quality, ground truth generation, computational method accuracy, and benchmarking rigor.*

In this article, we describe in detail the current computational challenges that limit the successful applications of deconvolution methods. We enlist the challenges in reference data, the lack of tools to generate realistic simulation data, the bias and accuracy issues in computational methods, and the demand for evolvable and reusable benchmarking studies. We conclude this report with a series of recommendations to solve these issues from the computational perspectives.

## Challenge 1: reference data quality

Deconvolution is critically dependent on the availability and accuracy of feature profiles for individual cell types in the form of a reference matrix S, where the rows represent the feature IDs (genes or DNA methylation sites) and columns represent the cell types. This reference matrix is typically supplied as an input to the deconvolution method, alongside the dataset from which cell type composition needs to be inferred. An ideal reference should contain all cell types present in the sample of interest, comprised of marker genes with large fold changes in relatively high and balanced numbers. Generating such reference matrices experimentally or computationally is challenging, due to the inherent complexity of cell types, the non-specificity of markers and the technical difficulty to dissociate them. Thus, in the initial stages and with the exception of blood there was an acute shortage of reference matrices. More recently, in the case of DNA methylation, reference matrices have been built for specific tissues including buccal swabs, saliva, cord blood, breast and pancreas, in most cases using sorted or cell-line data from existing resources such as ENCODE, the NIH Epigenomics Roadmap[11,12,13], or using in-house sorted cells[11,14].. 15Libraries of sorted cells have also been expanded for tissues like blood, allowing cell-type deconvolution at the resolution of 12 immune cell subtypes[16]. Using whole-genome bisulfite sequencing, a large DNA methylation atlas of sorted cell types encompassing most human tissues was recently generated[17], yet many key cell types remain missing (e.g. astrocytes, microglia in brain) and purity remains problematic. In the case of gene expression and spatial transcriptomics, the growing availability of tissue-specific scRNA-Seq atlases has allowed reference construction for most tissue types[18]. scRNA-Seq has also allowed the construction of DNA methylation reference matrices for effectively any tissue type using a probabilistic imputation approach[19,20]. However, in all cases, the quality of the constructed reference matrices remains a significant challenge. For instance, for single-cell based approaches cell-type annotation remains problematic and is still an active area of computational research. Single-cell assays may also miss specific cell types, or these cell types may be underrepresented. Some cell types, such as stem cells, do not have good consensus markers. Rare cell types are even less likely to be enriched experimentally. At higher resolution, also comes the difficulty of identifying cell-type specific markers with sufficiently large fold changes between similar cell types[21], leading to highly co-linear reference profiles and unstable downstream statistical inference.

The inconsistency among references for the same tissue is another common problem especially for gene expression data. Such inconsistency is due to a combination of many reasons, such as differences among biological samples (eg. patient age, gender, and ethnicity), geographic heterogeneity and sampling bias in bulk tissue specimens, environmental impact (eg. stress, environmental exposure, and smoking), sample preservation state (e.g., FFPE vs. fresh/frozen material), technical platforms (array vs RNA-Seq), experimental protocols (eg. sequencing depth that affects the detectability of certain genes), as well as the computational pipelines. As a result, the predicted fractions can differ substantially depending on which reference is used[22]. Even among the most widely referenced peripheral blood mononuclear cells (PBMCs), variations between reference datasets are observed for gene expression-based deconvolution[22]. Single cell RNA-Seq based reference approach can bypass some technical challenges due to sorted bulk cell references, but has its own set of issues as mentioned earlier, eg. missing cell types and differences in cell types due to annotation methods[23,24]. Furthermore, relative to scRNA-Seq, scDNA methylation experiments generate much sparser data, which limits the construction of reference cell type profiles.

## Challenge 2: ground truth data generation

Bulk transcriptomics data can be experimentally or computationally generated. Between them, the experimental approach undoubtedly resembles more the ground truth; however, the procedures to generate such data from solid tissue specimens are challenging. Some of these challenges are common to **Challenge 1**. Furthermore, one must be careful while experimentally deriving the "gold standard" for a cell type deconvolution task. For example, although the bulk transcriptomics data are usually obtained through the dissociation of solid intact tissue, the underlying cell-type proportions of this tissue are often orthogonally quantified on the basis of cell suspensions followed by assays such as scRNA-seq or FACS. Cell-type proportions are distorted in the cell suspension and do not necessarily reflect the proportions in the same solid tissue sample. As a result, assessing the performance of deconvolution methods by comparing the inferred cell-type proportions with the cell-type proportions measured from the single-cell assays will likely lead to misinterpretation during a benchmarking analysis. To ensure consistency, one possible solution is to dissociate the tissue specimen into the cell suspension, with part of the suspension used for bulk RNA-seq and part for single cell-based assays. Additional technical variations may also affect the evaluation of bulk sample deconvolution. For example, cryopreserved samples saved for different storage times may recover different "live" cell proportions, as compared to those obtained from fresh samples[25]. Some spatial transcriptomics studies have produced scRNA-Seq data from the same tissue sample to acquire the most relevant reference data to integrate with spatial transcriptomics[26,27]. Recent work mapped cells defined by spatial proteomics data to spatial transcriptomics data from adjacent sections in the same tissue block showing a higher accuracy than results from reference-based deconvolution method[28]. It is therefore pivotal to propose a better real-world solution to measure the ground truth, such as developing standards for paired bulk-seq and scRNA-seq to facilitate potential uses in cell type deconvolution.

Given these limitations, computational simulation of real-world tissue data may serve as a cost-effective, efficient, and useful way to provide various resources and benchmarking datasets. This approach is based on the assumption that the measured transcriptional patterns in a bulk sample are weighted linear combinations of reference profiles representing constituent cell types in the sample. The weights correspond to the relative proportions of cell types in the reference-based cellular deconvolution methods [29–31]. Therefore, in silico mixtures can be generated using purified cell-type-specific DNA methylation profiles or gene expression profiles with predefined cell type proportion weights [29,32]. Tools such as medepir [33] are available to generate simulated bulk DNA methylation profiles following this strategy. However, due to experimental difficulties in the purification and isolation of cell types (as stated in **Challenge 1**), as well as current limitations in single-cell DNA methylation references (e.g., data quality and cost), DNA methylation profiles for a number of cell types are missing in most solid tissues. This may skew benchmarking analyses toward certain tissue types, such as whole blood.

Alternatively, many computational tools are available for simulating scRNA-Seq count matrices, such as Splatter [34] and SymSim [35,36], which can be used to generate pre-designed and unbiased "ground truth". However, it is important to bear in mind that computationally generated scRNA-seq data for bulk mixture construction may not reflect or capture the full complexity from the real biological data, leading to overestimation of the real performance among the deconvolution approaches. Spatial transcriptomics data simulation has additional complexity from adding the spatial context, despite the relatively fewer cells and cell types contained in a spatial bulk sample (i.e., a spatial spot). A synthetic spatial dataset needs to capture the spatial distance between cell types, diverse mixtures of cell types across cellular neighborhoods, and spatial variation in gene expression and cell density across the tissue. Thus early attempts for creating synthetic spatial data have relied on scRNA-seq data or real spatial data to sample spatial gene expression values, but these values are then assigned to spatial coordinates without simulating a new spatial context [37–39].

## Challenge 3: limitations of computational methodologies

Despite the variations in omics data types, many issues related to deconvolution algorithms are shared among these omics and they have been addressed in prior benchmarking studies [40,41]. While the choice of the reference dataset (detailed in **Challenge 1)** is not directly part of the algorithms, it has been discussed by many studies to be the most important factor affecting deconvolution performance. Other issues include various aspects of "data pre-processing". For example, the data transformation and normalization steps affect the fitness of the input data for regression modeling, which assumes normal distributions of the input data. There is also collinearity among reference profiles, particularly on the cell types that are highly related (eg. CD4+ and CD8+ T cells; macrophages and monocytes). To deal with this, feature selection is often performed to maximize the robustness of the inference[42]. Additionally, the assumptions and suitable conditions are quite different in the reference-based vs. non-reference-based methods, explaining the difference in their performance and computational cost. In the following section, we individually discuss the computational methods based on the types of input data: bulk transcriptomics data, DNA methylation data, or spatial transcriptomic data.

**Transcriptomics-based deconvolution methods:** these methods can be classified into two main sub-categories, depending on the type of reference matrix data: bulk reference-based methods and single-cell reference-based methods.

Many bulk reference-based methods have been reported[22,40,43–47]. In the bulk reference-based methods, the reference matrix is usually generally derived from FACS-purified or in vitro differentiated cell subtypes[18]. Existing computational tools use various approaches to perform deconvolution. The most common one is regression, which can be further categorized as constrained linear regression approaches such as GEDIT[48],

DeconRNASeq[49], and quanTIseq [50], and non-constrained approaches such as CIBERSORTx[18] and DCQ [51]. Some methods can impute cell type-specific expression indirectly as well (e.g., DeMixT[52], CIBERSORTx[18], CODEFACS[53]). In particular, CIBERSORTx improves its performance by including different batch-correction modes to adjust variations between the single-cell reference and the bulk mixture. The dtangle[54] approach is different from other methods by modeling the linear mixing problem on a logarithmic scale. Scaden[55], on the other hand, is a neural network-based method trained on large-scale scRNA-Seq datasets to improve robustness. Additionally, many cell type quantification tools (eg. xCell) generate signature or enrichment scores for specific cell types among samples, rather than perform deconvolution explicitly[51,56–58]. These methods are limited when used to compare the proportions between different cell types, and they often underperform when repurposed for deconvolution[47].

Recently, new single-cell reference-based methods that use scRNA-Seq data to infer cell-type proportions in bulk samples have been developed[18,41,55,59–62]. Methods such as MuSiC[59] DWSL[61], and scDC[63] utilize single-cell references from multiple subjects to account for cross-subject differences and batch-effect confounding.

Multiple efforts to benchmark both types of transcriptomics-based deconvolution methods have been performed over the past several years[40,43–47,22,40,43–47]. Though this provides some guidance to users, the evaluated methods and reference data are highly variable (**Figure 2**). As a result, the consensus is lacking even among these benchmark studies (**Figure 2**), despite the fact that CIBERSORT/CIBERSORTx generally received a more positive recommendation, followed by MuSiC and EPIC. This leaves the question still open as to which tool(s) or methods should be used in specific conditions (see details in **Challenge 4**).

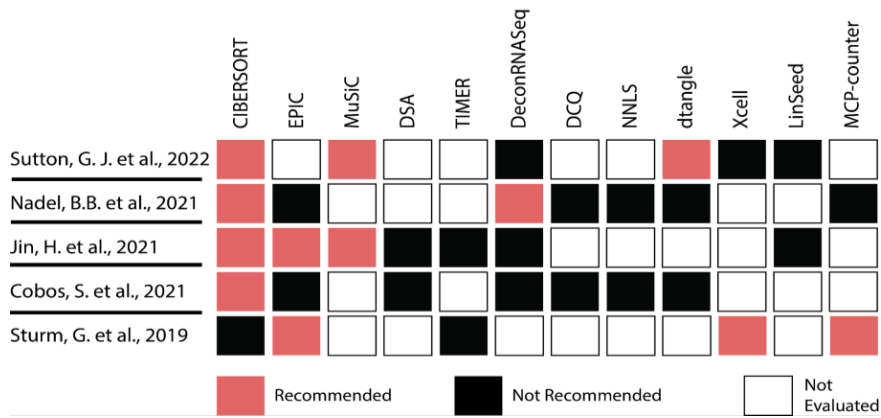

*Figure 2. Lack of agreement among different benchmark studies.* A heatmap showing recommendations from five benchmark studies[41,44,45,47] on selected bulk reference-based transcriptomic deconvolution methods (CIBERSORT[42], EPIC[64], DSA[65], TIMER[66], DeconRNASeq[49], DCQ[51], nnls[67], dtangle[54], xCell[58], LinSeed[68], and MCP-Counter[69]) as well as single-cell reference-based methods (MuSiC[59]). Only methods being evaluated by at least 2 benchmark studies are shown. The color scale represents different levels of recommendations for tools. Red: recommended; Black: not recommended; White: not evaluated.

**DNA methylation-based deconvolution methods**: Earlier computational deconvolution methods were more often designed for data generated from the array platforms (Illumina 450K or EPIC arrays). However, they also have technical challenges. The first one is to develop strategies that deal with the quality of reference data (**Challenge 1**). Newer statistical methods introduce various approaches to ensure the quality of the model or data. MethylResolver[32], a method based on least trimmed squares regression, allows the goodness of fit to be assessed. ARIC is another method based on a weighted support vector regression (SVR) to robustly estimate rare cell types[70]. EMeth detects and removes low-quality CpGs from the DNA methylation reference matrix, leading to improved inference[71]. To handle co-linearity among DNA methylation reference profiles, besides feature selection one can infer cell-type fractions hierarchically and iteratively at different levels of resolution, as implemented in the HEpiDISH algorithm[11]. This approach benefits from more flexible non-constrained inference approaches such as robust partial correlations (RPC)[11] or SVR which was originally implemented in CIBERSORT[42], where non-negativity and normalization constraints on the weights are imposed *a-posteriori*. An independent benchmarking study[29] showed that non-constrained methods (eg. RPC, SVR) can outperform constrained-based methods (eg. Houseman's Constrained Projection, or CP[72]). Besides the reference-based methods, reference-free methods such as RefFreeEWAS[73], BayesCCE[74] and TOAST[75], and semi-reference-free methods like RefFreeCellMix[76] are also available. RefFreeCellMix estimates cell proportion directly based on the methylation markers for each cell type, showed a promising improvement over the reference-free method. Multiple comprehensive benchmark studies have been done on array-based deconvolution methods[77,78,79]. The consensus is that reference-based methods produce higher specificity, but are less robust to confounding factors.

Most sequencing-based DNA methylation data, on the other hand, has much higher CpG site coverage compared to the array-based data. It thus requires the additional step of extracting the information from the selected regions, before cell type composition estimation. Multiple deconvolution methods are published, including MethylPurify[80], Bayesian epiallele detection (BED)[81], PRISM[82], csmFinder + coMethy[83], ClubCpG[84], and DXM[85], While BED and ClubCpG are reference-based, all other methods are reference-free. Some methods, such as DXM and PRISM, use Hidden Markov Models (HMM) to correct erroneous methylation sites before cell type proportion estimation. PRISM and MethyPurify use an EM algorithm to estimate the cell-type proportions. Interestingly, a recent benchmark study showed that the methods tailored for sequencing-based methods do not outperform the array-based methods such as Houseman's CP [86].

**Spatial transcriptomics-based deconvolution methods**: Spatial transcriptomics technologies enable the analysis of transcriptome information in the context of the tissue spatial organization. Among the various spatial transcriptomics platforms, NGS-based approaches[8,87–89] do not usually have the single-cell resolution in each spot. Thus, deconvolution is necessary for downstream cell-type proportion-dependent analysis for these technologies[8,90–124]. Similar to their non-spatial counterparts, many spatial transcriptomics deconvolution methods rely on the scRNA-Seq cell-type reference matrix, either from the same or different tissues[38,96]. Underlying methodologies for spatial transcriptomics deconvolution vary. Methods such as C-SIDE[125], RCTD[39], NMFReg[8], POLARIS[119], spatialDecon[120], and stereoscope [95] are based on regression modeling.

Methods such as DestVI [94], CellDART [99], GraphST[102], DSTG[97], Bulk2Space[100], Tangram[126], SD2[109], spSeudoMap[110], Antisplodge[104] employ deep learning models for deconvolution. Methods based on optimal transport such as SpaOTsc[114], and NovoSpaRc[113] can also be applied for the deconvolution purpose, although these methods were not specifically designed to address this task. Bayesian hierarchical models such as BayesTME[101], models making use of spatial information such as CARD[103], NMF methods such as SPOTlight[96] and NMFReg[8], convex optimization-based methods such as CytoSPACE[122] were also developed for deconvolving spatial transcriptomics data. Other methods include EnDecon[111] (ensemble learning), CellTrek[121] (random forest), STRIDE[127] (topic modeling) and methods originally developed for single cell annotation analyses such as Seurat[124] (canonical correlation analysis). With additional data dimensions, such as imaging and spatial distance, new deconvolution approaches for spatial data are emerging. For example, Tangram[126] can utilize the histology imaging data, if available, for spatial transcriptomics deconvolution. Furthermore, there are reference-free methods such as STdeconvolve[128] which uses a latent Dirichlet allocation model, and SPICEMIX[117] based on NMF. Similar to the recent trend in the deconvolution of methylation data, new semi-reference-free methods for spatial transcriptomics, e.g. Celloscope [115], allow for incorporating prior knowledge of gene markers for each cell type, while does not require an external single cell dataset.

There have been a few very recent benchmark studies on spatial transcriptomics deconvolution[129,130,131,132]. Among the commonly surveyed methods, again reference-based methods tend to do better than reference-free methods (**Figure 3**). More relevant references result in more accurate deconvolution. The consensus thus far is that cell2location[38], RCTD and stereoscope are the generally superior methods, followed by spatialDWLS[92]. However, other recommended methods vary and many existing deconvolution methods remain un-tested in the benchmark studies. Such inconsistency in the benchmark results is due to a combination of reasons, including different reference datasets, testing datasets, gold standards as well as evaluation metrics.

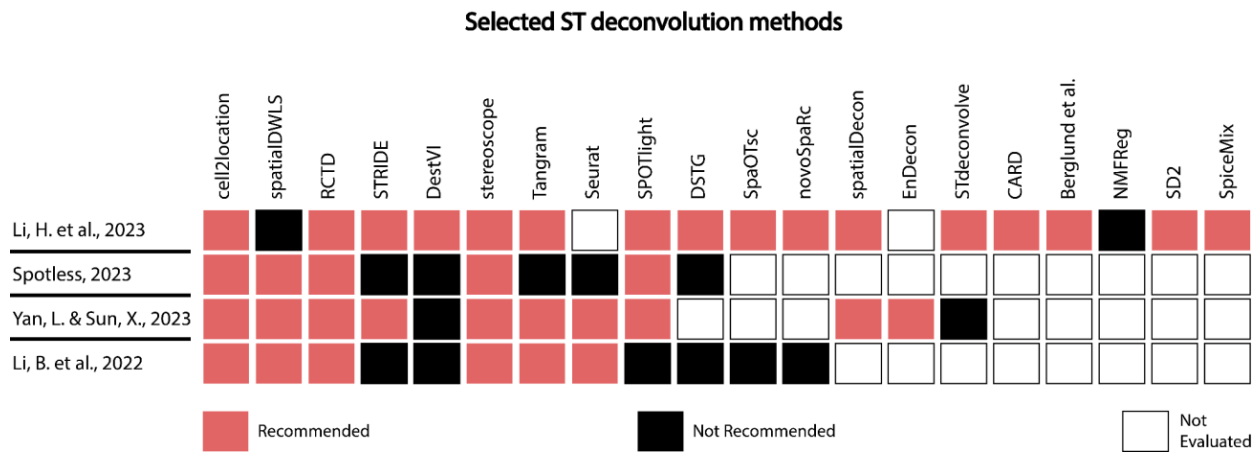

*Figure 3. Comparison among different spatial transcriptomics benchmark studies.* A heatmap showing method recommendations from four benchmark studies [129,130,131, 132]. Methods from left to right: cell2location[38], spatialDWLS[92], RCTD[39], STRIDE[127], DestVI[94], stereoscope [95], Seurat[124], SPOTlight[96],

*DSTG[97], SPOTlight[96], NovoSpaRc[113], spatialDecon[120], EnDecon[111], STdeconvolve[128], CARD[103], berglund[116], NMFReg[8], SD2[109], SPICEMIX[117]. The color scale represents different levels of recommendations for tools. Red: recommended; Black: not recommended; White: not evaluated.*

## Challenge 4: benchmarking design and implementation

Multiple benchmarking studies have been performed to guide users on tool selection, however, some of these studies give inconsistent recommendations despite evaluating the same tools [22,40,43–47]. The lack of consensus is probably due to a combination of reasons, including computational algorithms, reference data, as well as evaluation metrics.

Moreover, preprocessing steps on the input data, such as data transformation and normalization procedures, may also affect deconvolution performance. While the normalization method may significantly impact deconvolution performance for some methods such as EPIC[64], other methods such as CIBERSORT are more robust[43]. Unifying preprocessing steps required by different deconvolution methods can also be challenging. In addition, both simulated and real data have caveats as benchmarking datasets (see **Challenges 1 and 2**). Besides methods and data issues, the agreement is also lacking on the metrics used to evaluate deconvolution performances. Many benchmark studies[22,40,43,47] compute the Pearson correlation coefficient (PCC) for linear concordance and root-mean-square deviation (RMSE) for error measurement. However, not all benchmark studies have robustness measures that reflect deconvolution performances under different sets of deconvolution challenges. For example, the methods' ability to capture rare cell-type proportions was not commonly evaluated in most benchmark studies.

Perhaps the most significant issue above all is that these benchmark studies are not readily reusable[133]. Current benchmark pipelines are not designed to allow an easy plug-in comparison for new computational methods. Many do not offer the flexibility of user-selected input data and/or reference matrix, which impedes users' adaptation to benchmarking pipelines.

## Perspectives and recommendations

### Recommendation on high-quality reference data generation

One of the biggest challenges that computational deconvolution research faces is the availability of high-quality reference data from well-designed studies and biologically relevant samples. In the earlier phases of deconvolution methodological development, different computational methods utilized different reference datasets, and thus the performance differences among these methods may be largely driven by the quality of the reference dataset used, rather than the deconvolution algorithms themselves[43]. With the advancement of scRNA-Seq technology, tissue and organ level atlases are emerging as the new source of reference datasets. Although the measurable cell types are comprehensive, scRNA-Seq profile differences between studies are often observed and cell types are annotated differently by original studies. In the recently published Human Lung Cell Atlas[134], the authors integrated data of 2.4 million cells from 486 donors obtained from 49 datasets, this required extensive collection of metadata, developing methods for benchmarking data integration[135] and a combination of computational and manual methods for optimal cell annotations. Similar intense efforts will be required for every organ system to make sure quality and fidelity of reference data

Additionally, it is unclear which source of genomics data types serves as the best possible reference, putting aside other practical considerations of studies (eg. sample conditions). Is it scRNA-Seq data, bulk RNA-Seq data, bulk DNA methylation data or a combination of them? To answer this question, we will need community effort. One such representation is the tumor deconvolution DREAM challenge (projectID: syn15589870), which made a first attempt at creating standardized datasets and benchmarking standards for expression deconvolution[136]. More importantly, organization through consortiums will be pivotal, to generate various types of genomics reference datasets from the same tissues and samples using the standard experimental protocols during data generation. Such a painstaking initiative, focusing on controlling confounding factors and enabling cross-genomics evaluations, is not only necessary but critical for the computational deconvolution field to move forward. Fortunately, some communities and consortia have been established to accomplish part of the tasks, such as Human Cell Atlas (HCA), Human BioMolecular Atlas Program (HuBMAP), BRAIN Initiative Cell Census Network (BICCN) that aims to unify cell type nomenclatures and standards using the controlled vocabulary of the Cell Ontology[137]. Lastly, there are still some highly dynamic tissue types that are under-surveyed, such as human placentas[138]. Initiating new studies to obtain sufficient high-quality reference datasets will be instrumental to advance such areas of research.

Worth mentioning, compared to the cell type deconvolution - the focus of this article, deciphering the cellular states (or phenotypes) are even more complex and challenging as the differences between cell states are more subtle and the presence or absence of a cell state is dependent on biological conditions of even in the same organ or tissue, such as the immune cell states[139] . In addition to all the challenges related to cell-type annotation and cell-type specific reference profiling, one needs to consider the much larger feature space of the cellular states [134]. And as a result, specific repertoires of cell states are needed to successfully resolve this issue.

**Recommendation on new directions of computational methodologies**

With ongoing advances in single-cell genomics, computational methods using single-cell references may show advantages in increasing accuracy and robustness in deconvolution. Most transcriptomics- and DNA methylation-based deconvolution methods utilize traditional statistical methods to solve the linear mixing model of deconvolution. Given the vastly available large-scale omics data such as those generated from scRNA-seq, more robust methods such as neural-network-based models can be considered. Additionally, given that no universal best-performing method seems to exist for all data sets in each omics data type, ensemble approaches, such as EnsDeconv as reported recently [140], can be explored for general purposes. Probabilistic approaches have attempted to account for missing cell types in the reference dataset, suggesting unknown cell types in the query dataset that are not present in the reference [95,128]. Deconvolution methods on rare cell types and/or continuous sub-types are needed and should be assessed thoroughly with regards to detection sensitivity and robustness, regardless of RNA-Seq, DNA methylation or spatial data.

As mentioned earlier, due to the lack of high-quality and large-scale single-cell DNA methylation data, cell-type deconvolution of bulk-tissue DNA methylation data is an area needing more attention. Imputing DNA methylation reference matrices from corresponding gene-expression references derived from scRNA-Seq data, as implemented in EpiSCORE[19,20], is a promising approach but the imputation is promoter-centric and can only be carried out for 10-20% of expression marker genes. Thus, more inclusive and precise imputation methods are needed. For instance, imputation could include gene-associated enhancer regions, which should improve specificity among more similar cell types. Future efforts using imputation will be needed to improve the discriminating power among similar cell types. Alternatively, recent efforts to build a whole genome bisulfite sequencing (WGBS) based DNA methylation atlas of over 40 sorted cell types offered a more direct means to build DNA methylation reference matrices[17]. However, cell-type purity is not always high and for certain tissues, important cell types have not been profiled. Future computational work using this resource will thus require either using surrogate cell types from other related tissue types, or inferring them through AI-based learning processes.

Relative to bulk genomics data such as DNA methylation and gene expression, deconvolution methods for ST data are more complicated due to several reasons. The spatial neighborhood effect needs to be considered when deconvoluting spots, unlike gene expression or DNA methylation. Additionally, with fewer cells (10-20) and much more noise measured in each spot (as compared to bulk gene expression or DNA methylation), it is more challenging to accurately estimate cell-type proportions. Moreover, the technical variations across different cell types when measured by different technological platforms have not been quantified [141]. Lastly, To date most deconvolution methods that take into account platform effects assume the same effect size across different cell types per platform[39]. While it is unlikely that one single deconvolution method will be the best performing across all platforms, future choices of deconvolution methods need to consider generalization vs specialization on particular platforms. For example, when informing decisions on clinical samples, specialized methods will be optimal; whereas when data are obtained from heterogeneous platforms, then a generalized deconvolution method will be more appropriate.

**Recommendation on promoting rigorous benchmarking.**

Benchmark methods should incorporate sufficient flexibility to allow fair comparison, and at the same time restrict the variations in the pre-processing steps that interfere with the performance comparison. Furthermore, the evaluation metrics should reflect the main focus and the parameters studied, as well as the selection of benchmarking datasets. Other secondary measurements such as computational requirements, documentation quality, and installation instructions should also be evaluated from the user's perspective [142]. To be comprehensive, the benchmarking should include targeted datasets from different technological platforms (eg. both arrays and sequencing platforms). Cell type identification can also benefit from recent advances in single-cell multiomics and spatial multiomics, where integration of complementary data can enhance detection sensitivity and allows for cross-verification of cell types [143]. It should also consider real biological datasets as well as simulated data to draw consensus conclusions, as each of them has unique strengths and weaknesses. More realistic simulation approaches should be developed to capture the key characteristics of the data types. For example, the simulation of spatial transcriptomics should capture the spatial context such as diverse cell-type composition in different spatial neighborhoods and spatial changes in gene expression and cell density across the tissue. Simulation approaches could be benefited from advanced generative algorithms and by making use of high-quality benchmarking datasets that are available.

Benchmark studies also need to be forward-thinking, in addition to providing recommendations to both software developers and users. To enable better accessibility and allow the research community to build on existing evaluations as new methods are developed, it would be very helpful to present the data preprocessing, deconvolution method implementation, and metrics evaluation in a reproducible way (such as an R package or a conda environment). To allow for the inclusion of new methods in preexisting benchmark studies, software such as DeconBench[144], pipeComp [145] and CellBench [146] are ideal. For data dissemination, platforms such as Snakemake [147] or Docker (https://www.docker.com/) should be considered to package benchmark pipelines, so that the community can reuse the datasets and methods [148].

## Conclusions

Addressing intra-sample cellular heterogeneity is one of the most significant challenges faced by the genomics community. In this article, we discussed the major computational challenges in developing and applying deconvolution methods ranging from reference availability, data simulation, and computational methods to benchmarking evaluation. By raising awareness of those issues and proposing possible solutions, we expect to gather the community to push forward this area of research.

## Author's Contribution

LXG initiated and led the project. LG, YL and QH wrote the initial manuscript, YL and LXG made the figures with the help from QN. LXG, YL, CX, ST, NK, MP, AET and QN revised the manuscript.

## Acknowledgement

We would like to thank Drs. Brian Nadel, Saurabh Gupta, Serghei Mangul and Ankur Sharma for the discussions in the early stage of this project.